\documentclass[prl,
floatfix,
twocolumn,
superscriptaddress,
]{revtex4}

\usepackage{graphicx}
\usepackage{braket}

\graphicspath{{pics/}}

\begin{document}

\title{Fault-tolerant linear optical quantum computing with
small-amplitude coherent states}
\author{A. P. Lund} 
\email{lund@physics.uq.edu.au}
\author{T. C. Ralph} 
\affiliation{Centre for
Quantum Computer Technology, Department of Physics, University of
Queensland, St. Lucia, QLD 4072, Australia }

\author{H. L. Haselgrove} 
\affiliation{C3I Division, Defence Science and Technology Organisation, Canberra, ACT 2600, Australia}
\affiliation{School of Information Technology and
Electrical Engineering, University of New South Wales at ADFA,
Canberra 2600 Australia}

\begin{abstract}
Quantum computing using two optical coherent states as qubit basis
states has been suggested as an interesting alternative to single
photon optical quantum computing with lower physical resource
overheads. These proposals have been questioned as a practical way of
performing quantum computing in the short term due to the requirement
of generating fragile diagonal states with large coherent amplitudes.
Here we show that by using a fault-tolerant error correction scheme,
one need only use relatively small coherent state amplitudes ($\alpha
> 1.2$) to achieve universal quantum computing.  We study the effects
of small coherent state amplitude and photon loss on fault tolerance
within the error correction scheme using a Monte Carlo simulation and
show the quantity of resources used for the first level of encoding is
orders of magnitude lower than the best known single photon scheme.
\end{abstract}

\maketitle


Linear optical quantum computing uses off-line resource states, linear
optical processing and photon resolving detection to implement
universal quantum processing on optical quantum bits (qubits)
\cite{KOK}. This technique avoids a number of serious problems
associated with the use of in-line non-linearities for quantum
processing including their limited strength, loss, and inevitable
distortions of mode shape by the non-linear interaction. The trade-off
for adopting the linear approach has been large overheads in resource
states and operations. In the standard approach, which we will refer
to as LOQC \cite{KLM}, single photons are used as the physical
qubits. Although progress has been made in reducing the overheads
\cite{var}, for fault-tolerant operation they remain very
high~\cite{DHN}.

An alternative version of linear optical quantum computing, coherent
state quantum computing (CSQC)~\cite{ralph:catcomputing}, uses
coherent states for the qubit basis. This is an unusual approach as
the computational basis states are not energy eigenstates and are only
approximately orthogonal.  Previous work on CSQC has concentrated on
the regime where coherent states are relatively large ($\alpha > 2$)
and the orthogonality is practically zero.  It has been shown that
CSQC has resource-efficient gates~\cite{JEO}.

In this letter we show how to build non-deterministic CSQC gates for
arbitrary amplitude coherent states that are overhead-efficient and
(for $\alpha > 1.2$) can be used for fault-tolerant quantum
computation.  We estimate the fault-tolerant threshold for a situation
in which photon loss and gate non-determinism are the dominant sources
of error.  As our gates operate for any amplitude coherent states,
proof of principle experiments are possible using even smaller
amplitudes.  Given recent experimental progress in generating the
required diagonal resource states \cite{var2} we suggest that CSQC
should be considered a serious contender for optical quantum
processing.


For this paper we will use the CSQC qubit basis $
\ket{0}=\ket{\alpha},\ket{1}=\ket{-\alpha}
$ where $\ket{\alpha}$ describes a coherent state with (real)
amplitude $\alpha$ (i.e. $\hat{a}\ket{\alpha} = \alpha \ket{\alpha}$).
These states do not define a standard qubit basis for all $\alpha$ as
$\braket{-\alpha | \alpha} = e^{-2\alpha^2} \neq 0$, but for $\alpha >
2$ this overlap is practically zero~\cite{ralph:catcomputing}.  
A general CSQC single-qubit state 
is
\begin{equation}
\label{general-qubit}
  N_{\mu,\nu} (\alpha) \left( \mu \ket{\alpha} + \nu \ket{-\alpha} \right),
\end{equation}  
where $N_{\mu,\nu}(\alpha)$ normalises the state and depends on the
coefficients of the state.  
A special case is the diagonal states with $\mu = \pm \nu$ which can
be written as $\ket{\pm} = N_{1,\pm 1} (\alpha) \left( \ket{\alpha}
\pm \ket{-\alpha} \right)$.  These states form the resource used when
constructing CSQC gates using linear optics and photon detection.  The
diagonal state with a plus (resp. minus) sign has even (odd) symmetry
and only contains even (odd) Fock states.
%
%
%
%
%
This means that a diagonal (i.e. $X$-basis) measurement can be
performed by a photon counter and observing the parity.

\begin{figure}
\includegraphics[width=7.5cm]{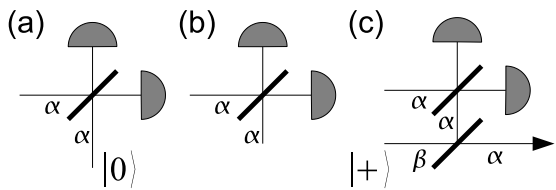}
\caption{\label{meas}Schematics for unambiguous CSQC (a) $Z$-basis and
(b) Bell state measurements and (c) CSQC teleportation.  Thin lines
represent modes whose state is a CSQC qubit with the encoding
amplitude shown near each line.  The $Z$-basis measurement in (a) as
described in~\cite{jeong:catcomputing} is performed by determining
which mode photons are present in.  The Bell state measurement in (b)
as described in~\cite{ralph:catcomputing} is performed by determining
which mode photons are in and how many photons are present.  Both
these measurements fail when no photons are present. (c) shows how
CSQC teleportation~\cite{bennett:teleportation,ralph:catcomputing} is
achieved. A Bell state is generated by splitting a $\beta = \sqrt{2}
\alpha$ diagonal state on a beam-splitter and performing a Bell state
measurement on an unknown qubit and one half of this entanglement.
All detectors are photon
counters, all beam-splitters are 50:50, and all unlabelled inputs are
arbitrary CSQC qubit states.}
\end{figure}

The computational or $Z$-basis measurement is shown in
FIG.~\ref{meas}(a) and the Bell state measurement is shown in
FIG.~\ref{meas}(b).  The $Z$-basis and Bell state
measurements must distinguish between non-orthogonal states.  For the
measurement to be unambiguous and error free it must have a
failure outcome~\cite{USD}.  This occurs in both measurements when no
photons are detected.  The probability of failure tends to zero as
$\alpha$ increases.


A critical part of constructing CSQC gates for all $\alpha$ is
teleportation~\cite{bennett:teleportation,ralph:catcomputing}.  This
is shown in FIG.~\ref{meas}(c).
As the teleporter uses unambiguous Bell state measurements there are 5
outcomes to the measurement.  Four outcomes correspond to successfully
identifying the respective Bell states.  When the appropriate Pauli
corrections are made the input qubit is successfully transferred
to the output.  The fifth outcome corresponds to the measurement
failure whose probability again decreases to zero as $\alpha$
increases.  Upon failure the output of the teleporter is unrelated to
the input and hence the qubit is erased.  It is this ability to
unambiguously teleport the qubit value, in spite of the fact that the
basis states are non-orthogonal, that is key to the success of our
scheme.


Unitary transformations on a CSQC qubit as defined in
Equation~(\ref{general-qubit}) will not reach all transformations
required to do quantum computing.  This is because unitary
transformations preserve inner products while various transformations
that we might wish to implement (e.g $\ket{\pm \alpha} \rightarrow
\ket{\alpha}\pm\ket{-\alpha}$) do not. 
We implement our gates using non-unitary,
measurement-induced gates which act like unitary gates on the {\em
coefficients} of our CSQC qubits for all $\alpha$.  This requires
gates which have in general a non-zero probability of failure.

We will construct a universal set of gates based
on~\cite{ralph:catcomputing} but applicable for all $\alpha$, that
allows us to implement error correction in a standard way.  Our
objective is to use the error correction to deal with gate failure
errors.

We will choose our universal set of quantum gates as a Pauli $X$ gate,
an arbitrary $Z$ rotation (i.e. $Z(\theta) = e^{i \frac{\theta}{2}
Z}$), a Hadamard gate and a controlled-Z gate.  Each gate acts on the
coefficients of the coherent state qubits as they would on orthogonal
qubits.

\begin{figure}
\includegraphics[width=7.5cm]{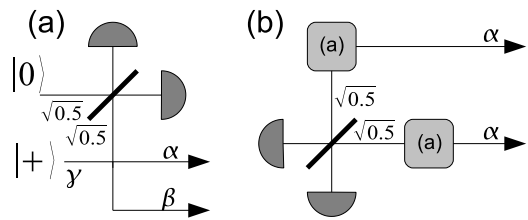}
\caption{\label{gates}Schematics for gate entanglement generation.
These diagrams have the same layout as those in FIG.~\ref{meas}. (a)
shows $Z$ rotation entanglement preparation.  A $\ket{+}$ state with
amplitude $\gamma = \sqrt{\alpha^2 + \beta^2 + 1/2}$ is split at a
three way beam-splitter generating the state
$\ket{\alpha^\prime,\alpha,\beta} +
\ket{-\alpha^\prime,-\alpha,-\beta}$ where $\alpha^\prime =
1/\sqrt{2}$ and $\beta=\alpha$ for the rotation.  The $\alpha^\prime$
mode is mixed at a beam-splitter with reflectivity $\cos \theta$ with
a coherent state of equal amplitude.  The two output modes are then
detected and the output is accepted if one photon is measured in total
(occurring approx. 1:3 times).
(b) shows the Hadamard entanglement preparation.  Two copies of the
entanglement from (a) are used but with different angles $\theta$ and
$\theta^\prime$ and one output mode with coherent state amplitude
$\beta = \sqrt{1/2}$.  Next, one $\beta$ mode from each
state are combined at a beam-splitter with reflectivity $\cos \delta$
and the output modes are detected.  The generation succeeds when only
one photon is detected in total.  If we choose the rotation angles as
$\theta = 3\pi/4, \theta^\prime = \pi/4, \delta = \pi/4$ and perform
an $X$ correction on one of the modes the desired entanglement is
produced. On average this procedure succeeds approx. 1:27 times.
}
\end{figure}

In CSQC the $X$ gate is the only gate deterministic for all $\alpha$.
The gate is performed by introducing a $\pi$ phase shift on the
qubit~\cite{ralph:catcomputing}.  The remainder of the gates are
implemented via quantum gate teleportation~\cite{gateteleportation}.
Just as we are able to implement unambiguous state teleportation, we
are able to implement unambiguous gate teleportation.  The gates are
implemented by altering the form of the entanglement used in the
teleporter.  The $Z$ rotation is achieved by using the entanglement
$
e^{i\theta}\ket{\alpha,\alpha} + e^{-i \theta}\ket{-\alpha,-\alpha},
$
the Hadamard gate uses the entanglement
$
\ket{\alpha,\alpha} + \ket{\alpha, -\alpha} + \ket{-\alpha,\alpha} -
\ket{-\alpha,-\alpha},
$
and the controlled-$Z$ uses the four qubit entanglement
\begin{eqnarray}
\lefteqn{\ket{\alpha,\alpha,\alpha,\alpha}
+\ket{\alpha,\alpha,-\alpha,-\alpha}} \nonumber \\
&  & +\ket{-\alpha,-\alpha,\alpha,\alpha}
-\ket{-\alpha,-\alpha,-\alpha,-\alpha}, \label{cz-ent}
\end{eqnarray}
which is used as the shared entanglement of two teleporters.  The
controlled-$Z$ entanglement can be generated from the Hadamard
entanglement with coherent state amplitude $\sqrt{2}\alpha$ by
splitting the outputs at 50:50 beam-splitters.  The procedures to
generate the Hadamard and $Z$-rotation entanglement  
are shown in
FIG.~\ref{gates}.



Depending on the outcome of the Bell state measurement in a teleported
gate, it may be necessary to apply an $X$ and/or $Z$ Pauli operator to
the output. 
In this paper, we assume that these Pauli operators are not applied directly,
but rather absorbed into the error-correction process via the {\em
Pauli frame} technique \cite{noisyknill}.  If the outcome of the Bell
state measurement is failure, then we say the gate failed and the
qubit on which it acted upon is erased.





In calculating a noise threshold for CSQC
it is necessary to establish a model for the noise experienced by each
operation (i.e. gates, measurements, and preparations). This model is
expressed in terms of two parameters: the qubit amplitude $\alpha$,
and a loss parameter $\eta$ (see below). We use this model to simulate
concatenated fault-tolerant error-correction protocols. A particular
setting of the parameters $(\alpha,\eta)$ is said to be {\em below the
threshold} if the rate of uncorrectable errors is observed to decrease
to zero as more levels of error correction are applied. Here we
calculate the {\em threshold curve}, defined to be the curve through
the $\alpha$-$\eta$ plane which lies at the boundary between the sets
of parameters that are above and below the threshold.


An important feature of our noise model is the inclusion of two types
of error: {\em unlocated} and {\em located} errors. A located error
occurs when a gate fails.  
The experimenter has knowledge about when and
where these errors occur. Unlocated errors are caused by photon loss
as these errors are not directly observable. Given that our noise
model includes both unlocated and located errors, 
we use an error-correction protocol which has been designed to deal
effectively with combinations of these two error types.  We have
chosen to utilise the ``circuit-based telecorrector protocol''
described in \cite{DHN}. This protocol uses error-location
information during ancilla-preparation and syndrome-decoding routines,
thus achieving a high tolerance to located noise, whilst achieving a
tolerance to unlocated noise similar to that of standard protocols
\cite{steane,reichardt}.

In practice, other noise sources would be present. Two examples are
mode mismatch and phase mismatch.  These noise sources will generate
additional unlocated and located errors in the teleported gates. The
effect of these errors will be similar to those in our simplified
noise model.  Depending on their strength, they may have a significant
effect on the noise threshold curve. We note that these errors are
{\em systematic}, and in principle can be greatly reduced by using
appropriate locking techniques.



The probability of gate failure varies as a function of
the input qubit state. 
For simplicity in the simulations, we apply the worst-case probability
value, which corresponds to the input state $\ket{+}$. The maximum
probability of failure (per qubit) for $Z$-basis measurements, and
Clifford group operations~\cite{cliffordgroup} implemented by gate
teleportation is equal to
%
\begin{equation}
q=\frac{2}{1+e^{2{\alpha^\prime}^2}}.
\end{equation}
In this equation $\alpha^\prime = (1-\eta) \alpha$ is an effective
encoding amplitude which incorporates the effects of loss. 
In the case of the controlled-$Z$ gate, this failure probability
applies independently to each of the two qubits.
%
%
Upon a gate failure the input qubit is erased.
For simplicity we model this effect by completely depolarising the
qubit upon a located error occurring.  



We model photon loss by assuming that each optical component, each
detector and each input coupling causes some fraction of the input
intensity to be lost, and that this loss is equal for all modes.  Due
to the properties of a linear network with loss it is possible to
assign one effective input coupling loss which incorporates all of
this loss together.  We also assume that the output of each gate
includes the loss due to the detectors from the {\em next} gate or
measurement.  From this we can assign an effective input loss rate
$\eta$ which combines the {\em detector, component and input}
efficiencies together incorporating all these effects.

The effect of loss on a CSQC qubit is to induce a random $Z$ operation
and decrease the coherent state amplitude~\cite{glancy:catloss}. We
assume that the decrease in amplitude is compensated by changing the
amplitudes of the coherent states in the entanglement used for the
teleported gates. The
%
%
probability of $Z$ error on a diagonal CSQC state is
\begin{equation}
p=\frac{1}{2}(1+\frac{\sinh{(2\eta - 1)\alpha^2}}{\sinh{\alpha^2}})
\end{equation}
where $\eta$ is the overall fractional loss as defined above. 

In the $Z$ rotation and the controlled-$Z$ gates, photon loss causes a
$Z$ error on the output state.  These are due to the loss in the
diagonal states from the generation of the entanglement.  In the
Hadamard gate, there are two diagonal states required and a loss in
one induces a $X$ error on the output and a loss on the other induces
a $Z$ error on the output (these errors are uncorrelated).

In our analysis we consider two noise models which are
summarised in TABLE~\ref{model}.
\begin{table}
\caption{\label{model}Error rates for the models used to calculate the
threshold curve for CSQC.  The coefficients in the $H$-gate and C-$Z$
gate arise from the larger $\alpha$ required for generating the
entanglement and are worse case.  Two models for qubit storage are
considered as shown in the row labelled ``Memory''.  In one model we
consider no noise in the operations that store CSQC qubits and the
second we introduce photon loss into these operations at the same rate
as introduced by the gates.}
\begin{ruledtabular}
\begin{tabular}{rccc}
& Loc. errors & Unloc. $X$ error & Unloc. $Z$ error \\ \hline
Memory & $0$ & $0$ & $p$ or $0$ \\
H-gate & $q$ & $1.6 p$ & $1.6 p$ \\
C-Z gate & $q$ & $0$ & $2.5 p$ \\
$\ket{+}$ & $0$ & $0$ & $p$ \\
X-meas & $0$ & $0$ & $0$
\end{tabular}
\end{ruledtabular}
\end{table}
%
%
%
%
%
%
We are considering here an
error-correction
protocol which consists of several levels of concatenation. The noise
model in TABLE~\ref{model} applies only to the lowest level of
concatenation (that is, to error-correction circuits that are built
using unencoded ``physical'' gates). For all higher levels of
concatenation, we assume a noise model identical to that considered
in~\cite{DHN} for the ``circuit-based telecorrection protocol'', since
the arguments used to derive that noise model are applicable to our
situation. Thus, our noise model and error-correction protocol are
identical to that of~\cite{DHN} for concatenation levels 2 and higher,
and so we do not perform new simulations for these concatenation
levels. Instead, we directly utilise the best-fit polynomials that
were obtained in~\cite{DHN}, in order to model the mapping between
noise rates and effective noise rates for all concatenation levels
other than the first.

\begin{figure}
\includegraphics[width=8.5cm]{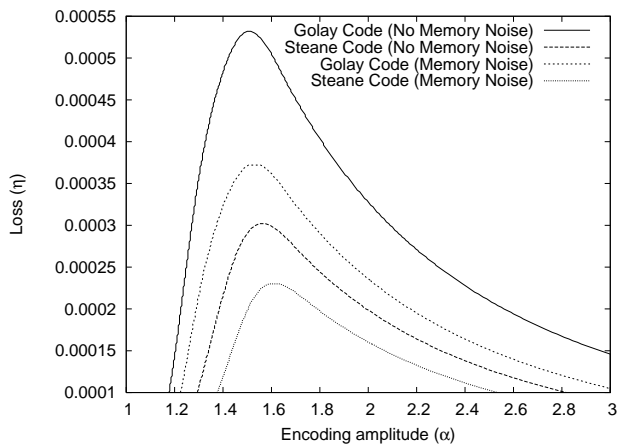}
\caption{\label{threshold-plot}Thresholds for CSQC using the 7 qubit
Steane and 23 qubit Golay code for both memory noise models.  
}
\end{figure}

%
%
%
For the first level of concatenation, we perform new numerical
simulations, for the noise models in TABLE~\ref{model}. The simulator
was a modified version of the one used in~\cite{DHN}.  All
controlled-\textsc{NOT} gates were replaced by controlled-$Z$ gates
and two Hadamard gates and simplifications of this circuit were
performed.
Separate simulations were performed for protocols based on the
$7$-qubit Steane code and the $23$-qubit Golay code. The resulting
threshold curves are shown in FIG.~\ref{threshold-plot}. An interesting
feature is that {\em increasing $\alpha$ beyond a certain point causes
a reduced tolerance to photon loss}.
%
%
%

%
TABLE~\ref{table1} estimates the resource-usage for one round of
error-correction, for 5 levels of concatenation.
\begin{table}
\caption{\label{table1}Effective error rates and resource usage for
the 7-qubit Steane code with memory noise enabled.  Coherent state
amplitude used for this table is $\alpha = 1.56$ and loss rate $\eta =
4 \times 10^{-4}$.  This corresponds to gate error rates in our model
of $(p,q)=(2\times10^{-4},0.015)$.  Resource usage is defined to be
the total number of gates, preparations, measurements, and quantum
memories used.  Resource are used in the following fractions for all
levels of concatenation: Memory 0.284, Hadamard 0.098, controlled-$Z$
0.343, Diagonal states 0.164, $X$-basis measurements 0.111. Also shown
is an estimate of the maximum length of computation possible assuming
the entire computation succeeds with probability $1/2$.}
\begin{ruledtabular}
\begin{tabular}{ccccc}
LEVEL & Unloc. & Loc. & 
Max. comp. & Resource \\ 
 & rate & rate & steps & usage \\ \hline
1 & $4 \times 10^{-4}$ & $8 \times 10^{-3}$ & $82$ & $1.0 \times 10^3$ \\
2 & $1.7 \times 10^{-4}$ & $2 \times 10^{-3}$ & $3.3 \times 10^2$ & $8.7 \times 10^5$ \\
3 & $2.8 \times 10^{-5}$ & $2.1 \times 10^{-4}$ & $3.0 \times 10^3$ & $4.5 \times 10^8$ \\
4 & $7.4 \times 10^{-7}$ & $3.6 \times 10^{-6}$ & $1.6 \times 10^5$ & $2.1 \times 10^{11}$ \\
5 & $5.3 \times 10^{-10}$ & $1.7 \times 10^{-9}$ & $3.1 \times 10^8$ & $9.6 \times 10^{13}$ 
\end{tabular}
\end{ruledtabular}
\end{table}
An advantage of CSQC over LOQC is lower resource usage. Using
TABLE~\ref{table1} and the success probabilities in
Fig.~\ref{gates} we find that CSQC consumes approximately $10^4$
diagonal resource states per error correction round at the first level
of concatenation. This is 4 orders of magnitude less than the number
of Bell pair resource states consumed under equivalent conditions by
the most efficient known LOQC scheme~\cite{DHN}. However, there is a
trade-off. The photon loss threshold we find for CSQC is an order of
magnitude smaller than that for LOQC. This means that if the loss
budget is too large then CSQC may not be scalable or may require so
many levels of concatenation that the resource advantage is lost.  We
note that the physical resources in terms of specific optical states
required to implement CSQC and LOQC are different.  Nevertheless we
believe comparing resource state counts still gives a good estimate of
the relative complexity of the two schemes.
%
%
%
In future work, it would be valuable to include other sources of
noise and improve upon some of the pessimistic assumptions made in
deriving the noise model.  One could consider ways of optimising the
fault-tolerant protocol in order to take advantage of the relative
abundance of $Z$-errors compared with $X$ errors.



We have shown how to construct a universal set of gates for coherent
state quantum computing for any coherent state amplitude.
Provided the coherent state amplitudes are not too small ($\alpha >
1.2$) and photon loss is not too large ($\eta < 5 \times 10^{-4}$) it
is possible to produce a scalable system. To our knowledge this is the
first estimation of a fault-tolerance threshold for non-orthogonal
qubits. As our gates work for all values of $\alpha$, proof of
principle experiments are possible using already demonstrated
technology.

We acknowledge the support of the Australian Research Council,
Queensland State Government and the Disruptive Technologies Office.


\end{document}